\documentclass[%
reprint, 
amsmath,amssymb,
aps,
prx,
  floatfix,
  longbibliography,
]{revtex4-2}

\usepackage{graphicx}%
\usepackage{dcolumn}%
\usepackage{bm}%
\usepackage{physics}
\usepackage[usenames, dvipsnames]{color}
\usepackage{sidecap}
\usepackage{textcomp}
\usepackage{textgreek}
\usepackage{amsmath}
\usepackage{xr}
\usepackage{minitoc}
\usepackage{siunitx}
\usepackage{bibunits}
\defaultbibliographystyle{apsrev4-2}
\defaultbibliography{bib1}

\definecolor{NewBlue}{rgb}{0, 0, 0.41}
\definecolor{NewRed}{rgb}{0.6, 0.07, 0.07}
\usepackage[colorlinks,
    linkcolor=NewBlue,
    citecolor=NewBlue,
    urlcolor=NewRed]{hyperref}

\DeclareSIUnit\angstrom{\text {Å}}
\renewcommand{\thesection}{\Roman{section}}
\newcommand{\suppsection}[1]{%
  \section{#1}%
  \addcontentsline{supp}{section}{\thesection.\quad #1}%
}

\newcommand{\suppsubsection}[1]{%
  \subsection{#1}%
  \addcontentsline{supp}{subsection}{\thesection.\arabic{subsection}\quad #1}%
}

\begin{document}

\title{High-fidelity two-qubit gates in a 7-qubit register for quantum networks}

\author{Margriet van Riggelen$^{1,2}$}
\thanks{These authors contributed equally to this work.}
\author{Jiwon Yun$^{1,2}$}
\thanks{These authors contributed equally to this work.}
\author{H. Benjamin van Ommen$^{1,2}$}
\author{Tim H. Taminiau$^{1,2}$}
\email{T.H.Taminiau@TUDelft.nl}

\affiliation{$^{1}$QuTech, Delft University of Technology, PO Box 5046, 2600 GA Delft, The Netherlands}
\affiliation{$^{2}$Kavli Institute of Nanoscience Delft, Delft University of Technology, PO Box 5046, 2600 GA Delft, The Netherlands}

\date{\today}

\begin{abstract}

Quantum networks based on optically active solid-state spins may enable quantum technologies including long-range quantum communication and distributed quantum computing. Network nodes containing multiple high-fidelity qubits can facilitate large-scale fault-tolerant operation. However, the stringent error thresholds remain out of reach for multi-qubit registers. In this work, we demonstrate high-fidelity two-qubit gates in a 7-qubit register, based on nuclear spins coupled to a nitrogen-vacancy (NV) center in diamond. We analyze crosstalk in highly connected spin systems, develop an efficient optimization procedure, and characterize the gates using gate set tomography. The two-qubit gate fidelities (best: 99.61(5)\%, average: 99.18(2)\%) demonstrate a multi-qubit register at the threshold for distributed quantum computation. Finally, as an example application, we perform a variational quantum eigensolver (VQE) simulation of the ground-state energy of H$_2$ and LiH molecules. These results demonstrate one of the key prerequisites for scalable quantum networks based on solid-state spins.

\end{abstract}

\maketitle
\newpage
\begin{bibunit}
\section*{Introduction}

Optically active solid-state spin qubits are a promising platform for future quantum technologies based on quantum networks, such as long-range quantum communication and distributed quantum computing and simulation \cite{Ruf2021, Wehner2018, Nickerson2013, Nickerson2014, DeBone2024, Singh2026}. In such a quantum network, multiple smaller registers (e.g. 3--20 qubits), called nodes, are linked together via optical interconnects. 
An optically active electron-spin qubit is used for internode entanglement and nuclear spins surrounding the color center are used as data qubits for memory and quantum information processing \cite{Knaut2024, Pompili2021, Stolk2024, Iuliano2026}. 

Based on these networks, large-scale distributed quantum computation has been proposed by spreading quantum error correction and logical qubits over the network \cite{Nickerson2013, Nickerson2014, DeBone2024, Main2024, Singh2026}. To satisfy the thresholds for fault-tolerant computation, high-fidelity quantum gates are required, in addition to fast high-quality optical links. For the distributed surface code with 3--6 qubits per node, detailed numerical simulations have found gate-fidelity thresholds starting from $\sim$99.4\% \cite{Nickerson2013, Nickerson2014, DeBone2024, Singh2026}, depending on the quality of the other operations, with various trade-offs and operation schemes possible. Despite recent progress \cite{Waldherr2014, Bradley2019, Beukers2025, Kuna2025, Song2025, Jaeger2026}, reaching such gate fidelities in multi-qubit systems remains a challenge for optically active solid-state spin qubits.

A promising candidate for high-fidelity multi-qubit control is the dynamically decoupled radiofrequency (DDRF) gate \cite{Bradley2019, VanOmmen2024, Kuna2025, Beukers2025}. This type of gate combines decoupling of the electron spin with radiofrequency (RF) driving of the nuclear spin to engineer a single- or two-qubit gate. In the original work \cite{Bradley2019}, no gate fidelities were measured, but rough estimates for two-qubit gate fidelities of 94-99\% were made by preparing Bell states. A more rigorous investigation and optimization of a DDRF two-qubit gate by Bartling et al. \cite{Bartling2024} recently reported 99.93(5)\% using gate set tomography (GST, \cite{Merkel2013, Nielsen2021}). That system, however, was limited to two qubits (the electron and nuclear spin of a nitrogen-vacancy center in isotopically purified diamond), which is not sufficient for distributed computation proposals \cite{Nickerson2013, Nickerson2014, DeBone2024, Singh2026}. Increasing the number of qubits introduces new challenges, most notably undesirable crosstalk between qubits \cite{VanOmmen2024} and the intractability of fully characterizing and optimizing gates in highly  connected multi-qubit systems.

In this work, we demonstrate two-qubit gate fidelities in a multi-qubit register at the threshold for scalable quantum networks. We control six nuclear spins coupled to an NV center in diamond with an average two-qubit DDRF gate fidelity of 99.18(2)\% (best gate 99.61(5)\%). We analyze the sources of decoherence and crosstalk in coupled electron-nuclear spin systems and develop a systematic and efficient optimization procedure to optimize DDRF gates. We then quantify the performance of the gates using two-qubit GST, a calibration-free characterization method that extracts the full gate process matrix and allows for distinction between different types of errors that occur during gates. As a testbed for the developed gates, we demonstrate variational quantum eigensolver (VQE) \cite{Peruzzo2014, McClean2016, Kandala2017, Guo2024} simulations of the ground-state energy of the molecules H$_2$ and LiH, in a hybrid classical-quantum algorithm that requires high-quality control of up to four qubits.

\begin{figure*}
    \includegraphics[width=1 \textwidth]{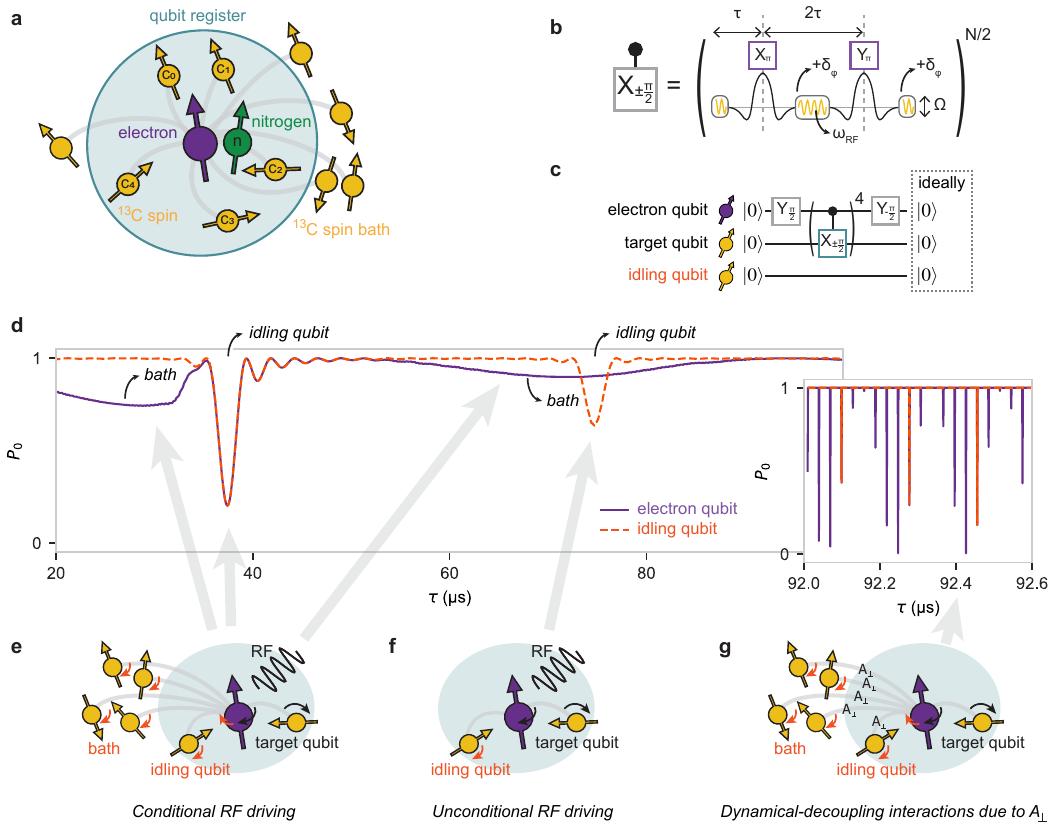}
    \caption{\emph{Crosstalk in a multi-qubit electron-nuclear spin system.} a) The 7-qubit register formed by the NV electron spin, the NV $^{14}N$ nuclear spin and 5 $^{13}C$ nuclear spins. Other spins in the $^{13}$C spin bath are not part of the register, but do cause decoherence of the register spins. b) Dynamically decoupled radiofrequency gates are used to engineer a two-qubit gate between the electron spin, and one target nuclear spin. The relevant gate parameters are $\tau$, $\omega_\text{RF}$, $\Omega$, $N$, and $\delta_\phi$. c) Quantum circuit to reveal the different types of crosstalk. The interpulse delay $\tau$ is varied while keeping the total rotation angle constant. After the circuit, all qubits should end in $|0\rangle$. d) Simulated outcome of the quantum circuit in (c). e-g) Diagrams depicting the different crosstalk mechanisms. Black (red) arrows represent intended (unintended) control. Crosstalk can arise through conditional RF driving of an idling qubit or the spin bath, leading to an electron phase-flip error and the idling qubit undergoing a bit-flip error (e) or unconditional RF driving of an idling qubit leading to idling-qubit bit-flip error (f). A third mechanism is resonant DD interactions mediated by $A_\perp$, causing electron-spin qubit phase-flip and/or idling-qubit bit-flip errors (g). }
    \label{fig:fig1}
\end{figure*}

\section*{Results}

\subsection*{System}

We consider a single NV center in diamond with a natural abundance (1.1\%) $^{13}$C concentration. We define a 7-qubit quantum register consisting of the NV electron spin ($m_s=\{0,-1\}$ subspace), the NV $^{14}$N nuclear spin ($m_I=\{0,-1\}$ subspace) and 5 $^{13}$C nuclear spins (Fig. \ref{fig:fig1}a). The $^{13}$C nuclear spins that are not included in the register form the $^{13}$C spin bath: we do not use them as qubits but they still play a role as a source of decoherence. We label the electron ``$e$'' and the nuclear-spin qubits ``$n$'' for the nitrogen and ``$c_0$", ``$c_1$" etc. for the $^{13}$C nuclear spins. The electron-spin qubit is controlled using microwave (MW) pulses ($\sim$GHz) (see Methods \ref{methods:setup}). The nuclear-spin qubits are manipulated using the DDRF gate \cite{Bradley2019, VanOmmen2024}.

\subsection*{DDRF gate parameter optimization}

A DDRF gate consists of RF ($\sim$MHz) pulses targeting a nuclear spin, which are interleaved with dynamical-decoupling (DD) pulses on the electron spin (Fig. \ref{fig:fig1}b). The five main parameters that define the gate are 1) the number of echo $\pi$-pulses $N$, 2) the interpulse delay $\tau$ and 3) the frequency $\omega_{RF}$, 4) the phases $\phi$ (given by a starting phase $\phi_0$ and a per-pulse phase increment $\delta_\phi$), and 5) the Rabi frequency $\Omega$ of the RF driving. The goal of this work is to develop the methods to optimize and characterize DDRF gates for all six nuclear-spin qubits. The target unitary for a two-qubit (or conditional) gate between the electron-spin qubit and a nuclear-spin qubit is $\ket{0}\bra{0}\ \otimes R_x(\frac{\pi}{2})+\ket{1}\bra{1}\ \otimes R_x(-\frac{\pi}{2})$, which is a CNOT gate up to single-qubit rotations. For the single-qubit (or unconditional) gate we target $\mathbb{I} \otimes R_x(\frac{\pi}{2})$ and $\mathbb{I} \otimes R_y(\frac{\pi}{2})$, which differ only by a $\pi/2$ shift in the starting phase $\phi_0$. For every gate, the objective is for the electron-spin qubit and a target nuclear-spin qubit to undergo the target unitary, without affecting the state of the other qubits in the register. Unwanted driving of other nuclear spins (crosstalk) can cause both phase flips of the electron-spin qubit and bit flips of the nuclear-spin qubits. In the following, we introduce the framework to construct a DDRF gate from the listed parameters, the mechanisms through which crosstalk could occur, and a DDRF gate optimization procedure.

\subsubsection*{Constructing a DDRF gate}

For the nuclear-spin rotation due to the RF pulses to coherently add up to $\pi/2$, a phase-resonance condition needs to be satisfied. The phase-resonance condition can be expressed in terms of the set RF phase increment $\delta_{\phi}$, applied before each RF element, 
\begin{equation}
    \delta_{\phi}=-(\Delta_0+\Delta_1)\tau + b\pi \text{ (mod }2\pi),
    \label{eq:resonance_condition}
\end{equation}
where $b\in\{0, 1\}$. $b=1$ results in opposite rotation axes of the gate for different initial electron-spin qubit states: a conditional or two-qubit gate. $b=0$ results in an unconditional or single-qubit gate on the nuclear-spin qubit. In this equation, $\Delta_0$ ($\Delta_1$) is defined as $\omega_{RF}-\omega_0$ ($\omega_{RF}-\omega_1$), where $\omega_0$ ($\omega_1$) is the precession frequency of the target nuclear spin when the electron-spin qubit is in state $|0\rangle$ ($|1\rangle$).

If this condition is met, the total rotation angle $\theta$ of the gate is given by  
\begin{equation}
    \theta = \tilde{\Omega}(\tau)N\tau.
    \label{eq:total_rot_angle}
\end{equation}
In this equation, $\tilde{\Omega}(\tau)$ is the \emph{effective} Rabi frequency, 
\begin{equation}
    \tilde{\Omega}(\tau) = \Omega(\text{sinc}(\Delta_1\tau) \hspace{4pt}\mp\hspace{4pt}\text{sinc}(\Delta_0\tau)),
    \label{eq:eff_rabi}
\end{equation}
where the sign corresponds to a conditional (-) or unconditional (+) gate \cite{VanOmmen2024}. 

Nuclear-spin qubit selectivity of the DDRF gate stems from both the bandwidth of the single RF pulse and from the phase-resonance condition in Eq. \ref{eq:resonance_condition} \cite{VanOmmen2024}. In this work, we choose $N=16$ or $N=32$ and $\tau>7$\textmu s. This ensures achievable Rabi frequencies allow for construction of a $\pi/2$-rotation and results in that the gate becomes long enough so that the selectivity comes mainly from the phase-resonance condition. This means that the oscillator frequency $\omega_{RF}$ plays little role in the quality of the gate. We will usually choose $\omega_{RF}$ such that $\Delta_1=0$. 

\subsubsection*{Crosstalk mechanisms}
Now we have the tools to investigate how crosstalk can cause electron-spin qubit phase-flips and idling nuclear-spin qubits' bit-flips. We focus on the impact of a nuclear-spin qubit $c'$ (the idling spin) while trying to perform a single- or two-qubit gate on a target spin $c$. The state on $c'$ should be unaffected, as should the action of the gate on $c$ and $e$. To realize the target gate, $\delta_\phi$ is set according to Eq. \ref{eq:resonance_condition}, $\theta=\pi/2$ (Eq. \ref{eq:total_rot_angle}), and $\Delta_1=0$. The circuit in Fig. \ref{fig:fig1}c, in which we set the electron-spin qubit in superposition and repeat the gate under study four times, can be used to expose different types of crosstalk that are discussed in this section.

The key lies in understanding that the $\delta_{\phi}$ for a conditional or unconditional gate on $c$ may simultaneously satisfy the resonance condition for a gate on $c'$:
\begin{equation}
\delta_{\phi}'
= -(\Delta_0'+\Delta_1')\tau +b'\pi \simeq \delta_{\phi},
\label{eq:crosstalk}
\end{equation}
where $\Delta_0'$ ($\Delta_1'$) is defined as $\omega_{RF}-\omega_0'$ ($\omega_{RF}-\omega_1'$), with $\omega_0'$ ($\omega_1'$) the precession frequency of $c'$ when the electron-spin qubit is in state $|0\rangle$ ($|1\rangle$). $b'$ indicates whether the interaction between the electron spin and idling spin is conditional or unconditional. How different $\delta_{\phi}'$ and $\delta_{\phi}$ need to be for crosstalk not to occur, depends on the duration of the gate: the longer the gate, the closer $\delta_{\phi}'$ and $\delta_{\phi}$ can be \cite{VanOmmen2024}. We first consider the case of an unintended two-qubit interaction between $e$ and $c'$ ($b'=1$). Assuming $\Delta_0 = \Delta_0'$ and using Eq. \ref{eq:resonance_condition}, we can rewrite Eq. \ref{eq:crosstalk} as follows:
\begin{equation}
\label{eq:condit_crosstalk}
(\Delta_1'-\Delta_1)\tau + b\pi = \pi \text{ (mod }2\pi).
\end{equation}
This type of conditional crosstalk leads to both a phase-flip error on the electron-spin qubit and a bit-flip error on $c'$. Since $\Delta_1'-\Delta_1$ does not depend on $\omega_{RF}$, \emph{whether} crosstalk can occur does not depend on the driving frequency. However, the \emph{amount} of crosstalk is determined by how much the idling qubit is driven. The idling qubit has its own effective Rabi frequency $\tilde{\Omega}’(\tau)$, given by Eq. \ref{eq:eff_rabi} for a conditional gate and dependent on $\Delta'_0$ and $\Delta'_1$. The conditional crosstalk described by Eq. \ref{eq:condit_crosstalk} results in rotation of the idling spin by angle $\theta' = \tilde{\Omega}'N\tau$ around some axis in the x-y plane, and a phase $\theta'$ acquired by the electron-spin qubit (Fig. \ref{fig:fig1}d). Because the electron is affected, this type of crosstalk is important to avoid even if the idling spin is not used as a qubit in the register. 

In addition to the conditional crosstalk, unconditional crosstalk may also occur, in which case the idling spin $c'$ is unintentionally driven, while the electron-spin qubit remains untouched (Fig. \ref{fig:fig1}e). Following a similar process as for conditional driving, now taking $b'=0$, this type of crosstalk happens when
\begin{equation}\label{phase_with_tau_uncon}
(\Delta_1'-\Delta_1)\tau + b\pi = 0 \text{ (mod }2\pi).
\end{equation}
in which case $c'$ is driven with effective Rabi frequency given by Eq. \ref{eq:eff_rabi} for an unconditional gate. This type of unconditional crosstalk is only relevant if $c'$ is used as a qubit in the register, as it does not result in phase errors on the electron-spin qubit.

The third type of crosstalk that should be avoided is caused by dynamical-decoupling interactions due to the perpendicular hyperfine interaction $A_{\perp}$ \cite{Taminiau2014} (Fig. \ref{fig:fig1}f). Here again, depending on $\tau$ it can happen that $c'$ is driven around an axis which is conditional or unconditional on the state of the electron-spin qubit:
\begin{equation}
\label{eq:dd}
\tau = \frac{k\pi}{\omega_0'+\omega_1'},
\end{equation}
where odd $k$ corresponds to a conditional rotation and even $k$ corresponds to an unconditional rotation.

The different mechanisms for crosstalk listed in this section limit the gate parameter sets that construct a good DDRF gate. We emphasize that for a single target gate on a qubit $c$, all the other nuclear-spin qubits are idling spins $c'$ that can cause crosstalk. In the case of conditional crosstalk, even nuclear spins that are not used as qubits cause crosstalk via phase-flips on the electron-spin qubit. In the following, we develop an experimental optimization procedure to navigate this parameter landscape.

\subsubsection*{Experimental optimization procedure}

\begin{figure*}
    \includegraphics[width=1 \textwidth]{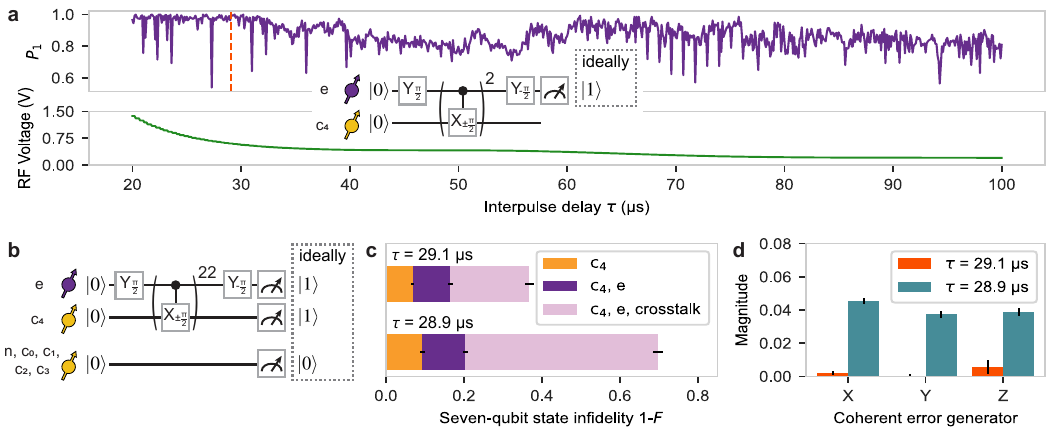}
    \caption{\emph{Measurements for the crosstalk during the $e$-$c_4$ two-qubit gate.} a) Upper panel: electron-spin qubit population in $\ket{1}$ after application of the circuit in the inset. Various regions and resonances where the coherence is reduced are observed. The red dotted line corresponds to $\tau=29.1$ \textmu s, the value we find for this gate at the end of the optimization procedure. Lower panel: RF voltage used to keep the gate rotation angle constant while sweeping $\tau$. b) Experimental circuit to amplify errors and detect crosstalk. Using the readout of all qubits, we compute the state infidelity after application of the circuit. c) State infidelity after application of the circuit in b). The orange bar indicates the state infidelity considering only the target qubit. Purple indicates the state infidelity taking into account both the target qubit and the electron qubit. The full pink bar indicates the seven-qubit state infidelity taking into account target qubit, electron-spin qubit and the other qubits in the register. c) Custom single-qubit GST experiment results. Red (Teal) bars correspond to the crosstalk gate on $c_4$ having interpulse delay $\tau=29.1$ \textmu s ($\tau=28.9$  \textmu s), which results in small (large) coherent error on target spin $c_0$. }
    \label{fig:fig2}
\end{figure*}

From Eqs. (\ref{eq:condit_crosstalk}), (\ref{phase_with_tau_uncon}) and (\ref{eq:dd}), it becomes clear that we should use $\tau$ to find a DDRF gate for which 1) we rotate the target nuclear-spin qubit by $\pi/2$, 2) we avoid conditional driving of both idling nuclear-spin qubits and nuclear spins in the spin bath, and 3) we avoid unconditional driving of idling nuclear-spin qubits. It is not experimentally feasible to do complete tomography of the 7-qubit register for every potential $\tau$ value that could result in a good gate. Therefore, we develop an optimization procedure to efficiently choose parameter values for all DDRF gates used in a multi-qubit register. 

In the first step, we explore a large range of $\tau$ values ($\sim$10-100 \textmu s) with a small amount of gates. For 40 promising values, we check for DD resonances nearby. We increase the number of gate applications as we reduce the number of promising $\tau$ values, based on electron and target spin readout. In the last step of the procedure, we probe crosstalk due to unconditional driving of nuclear-spin qubits. We do this by measuring the 7-qubit state fidelity after 22 applications of the gate, while the electron-spin qubit is in superposition. This indicates that the gate is expected to perform well in an algorithm that uses all of the qubits. More details about the procedure can be found in Methods \ref{methods:optimization}. 

We now discuss part of the optimization procedure steps for the two-qubit $e$-$c_4$ gate, as an example. Figure \ref{fig:fig2}a shows the first step of the procedure. We set the electron-spin qubit in superposition to be sensitive to phases originating from conditional rotations of the nuclear spins. We repeat the gate to be optimized twice and measure the electron-spin state (Fig. \ref{fig:fig2}a, upper panel). While we sweep $\tau$, we keep the rotation angle of the gate constant by compensating $\Omega$ according to Eq. \ref{eq:eff_rabi} (Fig. \ref{fig:fig2}a, lower panel, see Supplementary Information section \ref{suppsec:eff-rabi}). To choose promising $\tau$ values, we avoid regions and resonances where the coherence of the electron is reduced, because these are an indication of conditional crosstalk with the spin bath and individual spins respectively. Continuing the optimization procedure, we arrive at the final step with 5 candidate $\tau$ values. Only in the final step do we probe crosstalk due to unconditional driving, because it requires measuring all of the qubits, which is experimentally more expensive.  

The figure of merit we use to choose the final gate-parameter set is the 7-qubit state fidelity $F$ after application of the circuit shown in Fig. \ref{fig:fig2}b:
\begin{equation}
    F = F^{\ket{1}}_{\text{\text{target}}} \times F^{\ket{1}}_{\text{electron}}\times\prod_{i=\text{other qubits}}F^{\ket{0}}_i.
\end{equation}
While the two candidate values for $\tau$ that followed from previous steps in the procedure, $\tau = 29.1$ \textmu s and $\tau=28.9$ \textmu s, perform similarly well when only taking into account the electron-spin qubit and target qubit $c_4$, $\tau$=29.1 \textmu s performs better than 28.9 \textmu s when crosstalk is taken into account (Fig. \ref{fig:fig2}c). This is crucial for the usage of the gate in a multi-qubit algorithm.

To illustrate the importance of considering crosstalk, we compare the gate with $\tau=29.1$ \textmu s and the gate with $\tau=28.9$ \textmu s in a custom single-qubit GST experiment to demonstrate crosstalk (Fig. \ref{fig:fig2}d). GST aims to fit the measurement outcomes of diagnostic gate sequences to a self-consistent model that includes representations of state preparation, gates and measurements. This allows the extraction of the process matrix for each gate under test \cite{Nielsen2021}. To characterize crosstalk, we include an ``identity gate'' in a GST experiment on qubit $c_0$, which actually consists of four applications of the two-qubit gate on $c_4$. This gate ideally does nothing to $c_0$, the target qubit of the GST experiment, unless there is crosstalk. The coherent error generator values, that show unitary deviations of the gate from identity, should therefore be small. We see that this is indeed the case for $\tau=29.1$, while the increased values for $\tau=28.9$ \textmu s indicate crosstalk, as also signaled by the decrease in state fidelity.

\subsection*{Two-qubit gate set tomography}

\begin{figure*}
    \includegraphics[width=1 \textwidth]{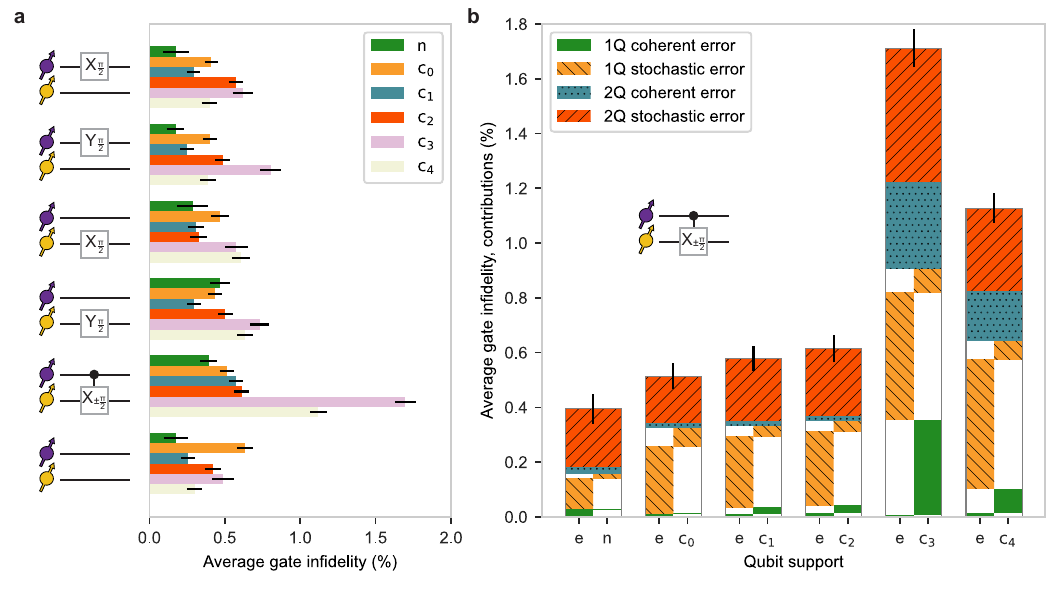}
    \caption{\emph{Characterization of the optimized gates} a) Average gate infidelity for the gates in our gate set. Data is obtained from six two-qubit gate set tomography (GST) experiments, each involving the electron-spin qubit (purple) and one nuclear-spin qubit (yellow). Error bars represent one standard deviation. b) We can separate different contributions to the average gate infidelity, here the two-qubit gates are shown as an example. Every column represents one two-qubit gate. Green and yellow half-bars indicate single-qubit errors on the electron-spin qubit (left) and nuclear-spin qubit (right), respectively. Teal and red full-width bars indicate errors on both qubits.}
    \label{fig:fig3}
\end{figure*}

We validate the optimization procedure using two-qubit GST, which has proven its versatility through application to various qubit platforms like phosphorus donors in silicon \cite{Madzik2022}, electrons in quantum dots \cite{Xue2022} and the NV center in diamond \cite{Bartling2024}. In this work, we use it to characterize 14 single- and 6 two-qubit gates, through 6 separate GST experiments. The gate parameters for these gates, found with the optimization procedure, are listed in Methods \ref{methods:gate_parameters}. In the previous section, we have focused on minimizing crosstalk due to unintended driving of idling spins. Now, we use GST to characterize the gates in the two-qubit space of the target qubits of the DDRF gate, to obtain the full electron-nuclear process matrices of the gates and to obtain performance metrics like average gate fidelity. 

The characterized two-qubit systems always include the electron-spin qubit and one nuclear-spin qubit. The gate set contains single-qubit $\pi/2$-rotations around the x- and y-axis for both qubits, a two-qubit gate, and an identity gate. The electron-qubit gates are defined including an XY4 decoupling sequence, to allow for any gate in the gate set to be followed by another gate in such a way that the electron spin is always protected by DD \cite{Bartling2024}. The single-qubit nuclear-spin gates and the two-qubit gate are the optimized DDRF gates. The identity gate is an XY4 sequence, to learn the coherent and stochastic error processes we have on the two qubits when we decouple the electron spin. This furthermore helps our understanding of the electron-spin qubit gate, because it allows for distinction between the error coming from the XY4 and the error coming from the MW pulse for the target operation. For more details on GST, see Methods \ref{methods:GST}.

Figure \ref{fig:fig3}a shows the average gate infidelity for all the characterized gates. For the single-qubit gates, we report the single-qubit gate infidelity in the full two-qubit space (see Supplementary Information section \ref{suppsec:on-target}). Notably, the average over the two-qubit gate fidelities on all six nuclear-spin qubits is 99.18(2)\% and for the $e$-$n$ two-qubit system, we find the best two-qubit gate fidelity of 99.61(5)\%. This result demonstrates a quantum network node with a qubit number and gate fidelities that are at or near the threshold for scalable distributed networks \cite{Nickerson2013, Nickerson2014, DeBone2024, Singh2026}.

In addition to the extraction of ``flat" performance metrics like average gate infidelity, GST gives information about different types of errors that occur during gates, such as coherent errors (e.g. overrotation) and stochastic errors (e.g. dephasing) \cite{Blume-Kohout2022}. We find that the fidelities of the electron-spin qubit gate and the XY4 identity gate are both mainly limited by stochastic phase errors on the nuclear-spin qubit. For the two-qubit gates, the contributions to the average gate infidelity of different error types is shown in Fig. \ref{fig:fig3}b \cite{Madzik2022}. The main error contribution to the one- and two-qubit DDRF gates is from stochastic phase errors on the electron-spin qubit. 

\subsection*{Variational Quantum Eigensolver}
Finally, we use the gates we have characterized and optimized to perform a testbed quantum algorithm: the variational quantum eigensolver (VQE). VQE is a promising future application of quantum simulation, that can be used to estimate the ground state energy of molecules \cite{Peruzzo2014}. VQE for molecular hydrogen (H$_2$), which requires just two qubits, has been investigated in various qubit platforms \cite{Peruzzo2014, OMalley2016, Kandala2017, Hempel2018, Ganzhorn2019,  Xue2022,  Guo2024}. Larger molecules (such as HeH$^+$, LiH, BeH$_2$, F$_2$) requiring control over more qubits, have only been explored in pioneering experiments with photons, superconducting qubits and trapped ions \cite{Peruzzo2014, Kandala2017, Shen2017, Hempel2018, Nam2020, Guo2024}. For optically active solid-state spin qubits, VQE for molecules has so far remained unexplored. The primary challenge is to combine high-fidelity multi-qubit control and efficient real-time classical-quantum optimization. In this work, we explore this algorithm for H$_2$ and lithium hydride (LiH). 

VQE is a hybrid quantum-classical algorithm, in which the quantum processor runs a quantum circuit (‘ansatz’) and outputs expectation values. These expectation values are then used by the classical computer to make a new suggestion for certain parameters in the circuit, based on a cost function. In this case, the cost function is the expectation value of the Hamiltonian $\left<H\right>$ of the molecule. Minimizing $\left<H\right>$ yields the ground-state energy of the molecule. 

The quality of the energy estimate is often evaluated using ‘chemical accuracy’. Chemical accuracy refers to a solution that is within 1.6 mHa of the exact solution, such that accurate predictions in quantum chemistry can be made \cite{Helgaker2002}. In the results discussed below, chemical accuracy is mostly not reached due to simplifications made in the ansatz circuit and decoherence and/or readout errors, as is the case in most previous experimental demonstrations \cite{Kandala2017, Hempel2018, Ganzhorn2019, Xue2022, Shen2017, Nam2020}. Chemical accuracy was reached for HeH+ using photons \cite{Peruzzo2014}, and using superconducting qubits for H$_2$ \cite{OMalley2016} and recently for LiH \cite{Guo2024}. 

Existing implementations differ in how they balance hardware feasibility against more expressive, complex quantum circuits (ansätze), which are generally more susceptible to gate and SPAM errors \cite{Guo2024, Hempel2018}. In this work, we adopt a hardware-efficient approach to constructing ansatz circuits, similar to that implemented for superconducting qubits by Kandala et al. \cite{Kandala2017}, which relies on real-time tuning of gate rotation angles. Unlike nuclear-spin qubit gates based on DD used in previous multi-qubit control \cite{Abobeih2022}, the DDRF gates used in this work naturally can be tuned to arbitrary gate angles by making use of the precision on the amplitude of the RF driving. We combine the hardware-efficient ansatz with a promising error mitigation technique proposed by Czarnik et al. \cite{Czarnik2021} and applied to VQE by Guo et al. \cite{Guo2024}, called Clifford fitting (CF).

\subsubsection*{VQE implementation}
We map the fermionic Hamiltonian for H$_2$ (LiH) to 2 (4) qubits (see Methods \ref{methods:VQE}) and implement hardware-efficient ansatz circuits, which consist of circuit layers of single-qubit gates, alternated with entangling layers \cite{Kandala2017}. Single-qubit gate $i$ implements the unitary $U_i(\boldsymbol{\theta}_i^k)$, defined by the Euler angles $\boldsymbol{\theta}^k_i$, which are varied every iteration $k$ of the algorithm to minimize $\left<H\right>$. For the electron qubit, $U_i(\boldsymbol{\theta}_i^k)$ is deconstructed into rotations around fixed axes as $Z_{\theta^k_{i,1}}X_{\theta^k_{i,2}}Z_{\theta^k_{i,3}}$, while for the nuclear-spin qubits we adjust the amplitude and RF phases of a single DDRF gate (see Supplementary Information section \ref{sec:parametrized_gates}). The initial parameter values $\boldsymbol{\theta}^0_i$ are chosen such that we start the optimization process by preparing the Hartree-Fock state. 

To minimize $\left<H\right>$, we employ a gradient-descent method called simultaneous perturbation stochastic approximation (SPSA) \cite{Kandala2017, Spall1992}. Every iteration $k$ of the algorithm $\left<H\right>$ is evaluated for two trial parameter sets, from which a gradient is calculated to determine $\boldsymbol{\theta}^{k+1}_i$ (see Supplementary Information section \ref{suppsec:vqe-implementation}). After 100 (50) trials for H$_2$ (LiH), we stop the optimization and measure $\left<H\right>$ for the best $\boldsymbol{\theta}^{k}_i$ with more experimental shots to decrease shot noise and obtain a final estimate for the ground state energy. The entire optimization procedure is performed for various spacings $R$ between the atoms, which correspond to different Hamiltonians $H$, to retrieve the potential landscape. Furthermore, we perform the experiment with and without CF. For CF, we run additional versions of the VQE circuits prior to the VQE optimization loop, in which most parametrized gates are replaced by Clifford gates. This way, we obtain a correction of the combined errors in our gates and SPAM for every Pauli-operator basis (see Supplementary Information section \ref{CF}). The correction is applied at every iteration of the VQE algorithm, before feeding the measurement outcomes to the optimization algorithm. 

\subsubsection*{Results H$_2$}

\begin{figure*}
    \includegraphics[width=1 \textwidth]{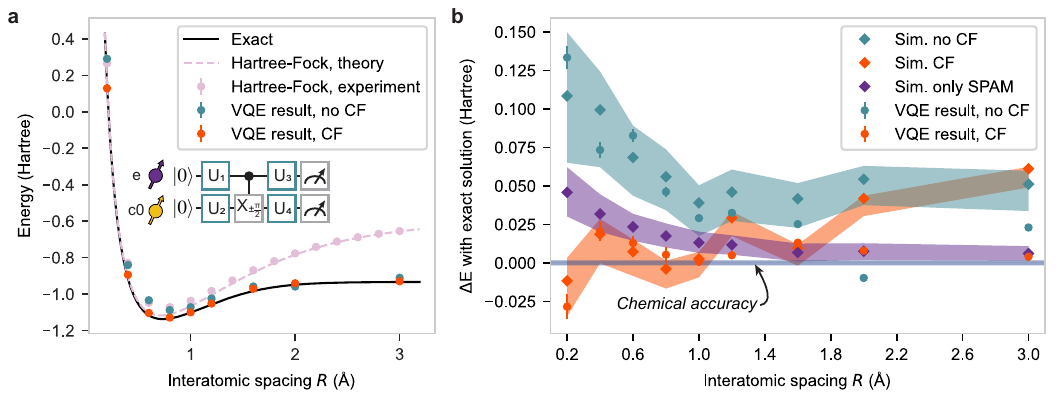}
    \caption{\emph{VQE for H$_{2}$} a) Ground state energy of H$_{2}$ for varied interatomic spacing $R$. Pink dots indicate the energy when constructing a HF state with the circuit, which can be compared to the theoretical HF energy value (pink dotted line). Teal (orange) dots indicate experimental VQE results without (with) clifford fitting (CF). The exact value is shown for comparison. Inset: two-qubit quantum circuit to calculate the H$_{2}$ ground state. b) Difference with the exact ground-state energy. Dots are the experimental VQE data, diamonds indicate the result of simulations based on two-qubit GST results. Shaded region is a 2$\sigma$ confidence region around the simulated values based on a Monte-Carlo simulation of the experimental shot noise. Purple diamonds indicate a simulated VQE experiment with SPAM error estimates taken from GST and perfect gates.}
    \label{fig:fig4}
\end{figure*}

The 2-qubit circuit for the $\mathrm{H}_2$ ground-state energy estimation consists of a two-qubit gate, surrounded by single-qubit gates on both qubits (Fig. \ref{fig:fig4}a). First, we use the circuit to construct the classical Hartree-Fock (HF) state. There is a discrepancy between the experimental and theoretical HF energy for small $R$, which is caused by a combination of gate and SPAM errors. Running the VQE algorithm starting from the HF state results in an improvement for almost all $R$, and we retrieve the shape of the potential landscape. CF improves the estimate of the ground state energy further for all $R$. Looking at the difference with the exact solution (Fig. \ref{fig:fig4}b), we see that chemical accuracy is achieved for one spacing. We compare this result with two different types of simulation.

The GST characterization of the two-qubit gates enables us to numerically simulate the VQE experiment. First, we verify that the information obtained with GST is useful to predict the outcome of the VQE experiment. To simulate the two-qubit gate, we take the characterized process matrix directly from GST. Also SPAM errors can be taken directly from the GST result, in the form of a density matrix for the initial state and a positive operator-valued measure (POVM) for the measurement. Because the single-qubit gates are different every iteration, we do not have their full action characterized. For the simulation, we separate the stochastic error of the electron-spin (nuclear-spin) qubit $X_{\frac{\pi}{2}}$ gate process matrix, and use it to simulate the error of the electron-spin qubit (nuclear-spin qubit) gates in the VQE circuit. Note that we don't have to run the optimization loop in simulation, instead we directly take the best $\boldsymbol{\theta}^{k}_i$ from the VQE experiment. To account for shot noise in the experiment, we use a Monte Carlo simulation to find a confidence region of two standard deviations. For small $R$, we find that both the results without and with CF are predicted accurately (Fig. \ref{fig:fig4}b). For large $R$, the VQE experiment performs better than predicted by GST. This could be due to the fact that we do not take coherent errors of the parametrized gates into account for the simulation, and that these errors can be exploited to minimize $\left<H\right>$ due to the variational nature of the algorithm.

In a second simulation, we separate different noise sources. We run a simulation with the SPAM errors present, but with perfect gates, shown in purple in Fig. \ref{fig:fig4}b. For this simulation, we do implement the optimization loop and we do not perform CF. We find that 1) gate errors significantly reduce the accuracy of the result and 2) even with perfect gates, we cannot reach chemical accuracy due to the SPAM errors. Comparing this simulation with the experimental result where we use CF, we can see that CF helps to mitigate both gate and SPAM errors. 

\subsubsection*{Results LiH}

\begin{figure}
    \includegraphics[width=0.5 \textwidth]{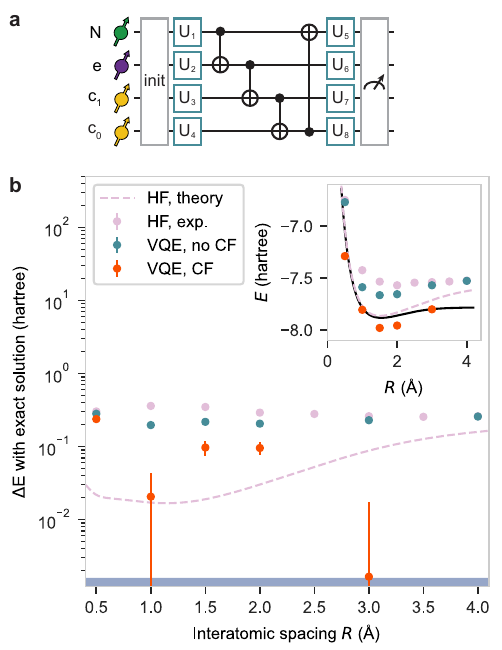}
    \caption{\emph{VQE for LiH} a) Quantum circuit for LiH VQE. b) Difference with the exact ground-state energy of LiH for varied interatomic spacing $R$. Pink dots are an experimentally obtained ground-state energy estimate by constructing the HF state using the VQE circuit. Teal (orange) dots dots indicate experimental VQE results without (with) Clifford fitting (CF). The exact values and the theoretical HF values are shown for comparison. Inset: Ground state energy of LiH.}
    \label{fig:fig5}
\end{figure}
To benchmark the gates in a multi-qubit context, we demonstrate VQE for LiH, which was previously implementing using superconducting qubits and trapped ions \cite{Kandala2017, Hempel2018, Guo2024}. The ansatz is compiled as a quantum circuit with four qubits and a fifth qubit is involved in the readout of the electron-spin qubit (see Supplementary Information section \ref{suppsec:init_and_ro}). We choose a cyclic two-qubit gate layer as the starting point for the hardware-efficient ansatz circuit (Fig. \ref{fig:fig5}a). While this `depth-1' circuit is too shallow to obtain a ground-state energy within chemical accuracy (see Supplementary Information section \ref{suppsec:lih_depth}), it is an interesting test-bed for our quantum gates, demonstrating hybrid quantum-classical control over 5 qubits. 

We compile the circuit into native operations. A prominent source of error in deeper circuits with more qubits is nuclear-spin dephasing. Therefore we integrate nuclear-spin qubit echoes in the circuit \cite{Abobeih2022} and maximize the time the nuclear-spin qubits are in an eigenstate (see Supplementary Information section \ref{suppsec:lih_circuit}). As a trade-off, integrating echo pulses takes the total circuit duration from 8.15 ms to 12.97 ms. While the nuclear-spin qubit coherence is extended, the electron-spin qubit now needs to maintain coherence for a longer time.

Figure \ref{fig:fig5}b shows the LiH VQE results for various interatomic spacings. Performing the VQE optimization improves upon the initial state (the HF state). The experimental HF results differ from the theoretical values, which is an indication that in this circuit the combination of SPAM and gate errors is significant. Additionally, we perform VQE with CF. CF brings the estimate of the ground-state energy consistently closer to the exact solution. Notably, using CF resulted in ground-state energies below the exact solution for all spacings (Fig. \ref{fig:fig5}c). We attribute this to fit errors in the CF process that are exploited by the VQE optimization algorithm to minimize the ground-state energy.

\section*{Discussion}

We have demonstrated two-qubit gate fidelities of, on average, 99.18(2)\% over a 7-qubit system, and implemented a VQE algorithm as a testbed for multi-qubit control. The gate fidelities are enabled by a systematic analysis of crosstalk errors and a novel gate optimization procedure. These insights can be applied to other optically active solid-state spin qubits, like group-IV defects in diamond \cite{Beukers2025} and defects in silicon carbide \cite{Kuna2025}. 

Our results realize an optically accessible 5-qubit register (qubits: $e$, $n$, $c_0$, $c_1$, $c_2$) with gate fidelities at the threshold for distributed fault-tolerant computation, assuming errors during other components, such as measurements and entanglement generation, are small \cite{Nickerson2013, Nickerson2014, DeBone2024, Singh2026}. Exact thresholds will depend on the parameters of all operations in the network, with a large number of trade-offs and different protocols possible \cite{Nickerson2013, Nickerson2014, DeBone2024, Singh2026}. Additionally, effective computation will require gate fidelities well above the threshold. Therefore, it is clear that future improvements in qubit control are still required.

Potential paths to explore are reducing the electron- and nuclear-spin-bath noise, while optimizing qubit-qubit interaction strengths, through (isotopic) material engineering \cite{Bartling2024, Yamamoto2026}, and mitigating decoherence through parallelization of DDRF gates to reduce sequence times. Combined with further theoretical advances in distributed error-correction codes \cite{Singh2026} and improved (distillation) protocols and decoders \cite{DeBone2024}, and experimental improvements of the spin-photon interface through optical cavities, integrated optics and novel color centers \cite{Kuruma2021, Knaut2024, Pasini2024, Codreanu2025}, our results form one of the prerequisites towards network-based distributed quantum computing based on solid-state color centers.

\section*{Methods}

\subsection{Setup and sample}
\label{methods:setup}
Measurements were performed on an NV center at cryogenic temperatures (4K, Montana Cryostation S50), in a diamond with a natural abundance of $^{13}\mathrm{C}$. The NV center is oriented along the $\left<111\right>$ axis, parallel to the surface-normal of the diamond. A solid immersion lens (SIL) is made around the NV center to improve the collection efficiency, and a gold stripline running along the SIL enables electron- and nuclear-spin control.

SPAM of the electron-spin qubit is performed by resonant optical excitation of spin selective transitions \cite{Robledo2011}, resulting in a single shot readout (SSRO) fidelity of $\sim$93\%. The measurements in the optimization procedure and for VQE are corrected for the SSRO infidelity. Prior to every experiment a charge-resonance check is performed to confirm the right NV charge state and optical resonance \cite{Bernien2013}. Nuclear spins are initialized by optically initializing the electron-spin qubit and swapping that state to an uninitialized nuclear-spin qubit. Nuclear spin readout is achieved in a similar fashion, by mapping the $Z$ projection of the nuclear-spin qubit state to the electron-spin qubit, and reading out the electron-spin qubit. For more details on initialization and readout see Supplementary Information section V. 

The magnetic field is applied with two permanent magnets mounted directly behind the diamond, inside the cryostat. External magnets are used to align the magnetic field along the NV axis. Every 30 minutes, the magnetic field is adjusted by moving the external magnets, to set the NV's $m_s=-1\rightarrow m_s=0$ transition frequency to 4.18683 GHz ($\sim 252.2$ mT), within a tolerance of 20 kHz. The microscope position is optimized every few hours to stabilize the photon collection efficiency.

The control sequences for the qubits are generated using a Zurich Instruments HDAWG. Microwave (MW) control of the electron-spin qubit (GHz) is performed using single-sideband modulation (SSB) using a Rohde \& Schwarz SGS100A microwave source, followed by amplification to 20W peak power, and a switch to minimize noise from the amplifier affecting the qubits \cite{Abobeih2018}. A second MW signal generator was used to address the $m_s=0\rightarrow m_s=+1$ transition of the electron spin. Hermite-shaped microwave pulses (duration 140 ns) are used to rotate the electron spin independently of the nuclear-spin states \cite{Bartling2024}. RF pulses on the nuclear spins (MHz) are generated using the HDAWG by modulating pulse envelopes with an internal numerical oscillator, enabling runtime adjustments of the RF phase to track the evolution of the nuclear spins. The RF pulses are amplified using an RF amplifier (Analog Devices ADA4870). To prevent the electron spin from gaining a phase due to the RF pulses, the length of an RF pulse is always $\tau^{RF}=n/f_{RF}$, with $n\in\mathbf{N}$ and $f_{RF}=\omega_{RF}/2\pi$ the pulse frequency. RF pulses are shaped with a $\sin^2$ roll-on and roll-off, which is always $t_\mathrm{roll}=4/f_{rf}$ long. This envelope causes the RF pulses with a duration of $\tau^{RF}$ to cause slightly less than half the rotation of the RF pulses with a duration of $2\tau^{RF}$, which is compensated by increasing the amplitude of the pulse of duration $\tau^{RF}$.

All data was taken using the QMI Python package \cite{Ervasti2026}.

\subsection{Gate optimization procedure}
\label{methods:optimization}
To construct a good DDRF gate in a multi-qubit register, we follow a fixed procedure consisting of four steps. Throughout the procedure we reduce the number of candidate values for interpulse delay $\tau$, which is the parameter that we use to explore the parameter space. We start with a calibrated reference gate.

In the first step, we set the electron in superposition, repeat the gate two times and return the electron to an eigenstate. We sweep $\tau$ between a lower bound and 100 \textmu s, in steps of 0.1 \textmu s. The lower bound for $\tau$ is determined by the maximum RF voltage. The estimated required RF voltage for the gate is determined by calculating the effective Rabi frequency \cite{VanOmmen2024}, see Supplementary Information section I. We measure the electron-spin state. Because we keep the total rotation angle to be $\pi$, any reduction in the coherence of the electron-spin indicates either crosstalk or decoherence. We choose the 40 values of $\tau$ that have the best coherence.

The goal of the second step is to discard $\tau$ values for which there is a DD interaction due to the $A_{\perp}$ of idling spins. To achieve this, we apply 128 DD pulses with the electron in superposition. To account for small magnetic field fluctuations, we do the measurement for the 40 selected $\tau$ values and additionally for values $\delta t =$417 ps smaller and larger than the values in the selection. We take the electron coherence average of the three points $\tau$, $\tau-\delta t$ and $\tau+\delta t$ and discard the worst 15.9\% of the values (six values). 

Now left with 34 potential values for $\tau$, we apply the gate 10 times, sweep the RF voltage, and readout the electron. This gives 5 $\tau$-voltage combinations for which we do the final step.

In the final step (Fig. \ref{fig:fig2}b, c), we apply the gate 22 times. We readout the electron-spin qubit, the target nuclear-spin qubit and one of the other nuclear-spin qubits (idling spin). We repeat the measurement for all the nuclear-spin qubits in the register as the idling spin. We choose the $\tau$ value that results in the best state fidelity for the electron-spin qubit, for the target nuclear-spin qubit and for the idling spins together. This ensures good performance in a quantum algorithm involving multiple qubits.  

\subsection{Gate parameters}

\begin{table}[h]

\begin{tabular}{c|c|c|c|c|c}
 & \shortstack{\phantom{g}$\omega_{RF}/2\pi$\phantom{g} \\ (kHz)} 
 & \shortstack{\phantom{g}$N$\phantom{g}\\ \phantom{(N)}}
 & \shortstack{RF voltage \\ (V)} 
 & \shortstack{\phantom{g}$\tau$ \phantom{g}\\ (\textmu s)} 
 & \shortstack{Total gate time \\ (\textmu s)} \\ \hline\hline
$n$  & 4173.388 & 32 & 0.5062  & 12.96 & 829.44  \\ \hline
$c_0$ & 2797.69  & 16 & 1.1597  & 16.7  & 534.4   \\ \hline
$c_1$ & 2754.55  & 16 & 1.4658  & 12.6  & 403.2   \\ \hline
$c_2$ & 2653.001 & 16 & 1.9345  & 12.4  & 396.8   \\ \hline
$c_3$ & 2680.491 & 16 & 0.47539 & 35.2  & 1126.4  \\ \hline
$c_4$ & 2676.095 & 16 & 0.687   & 29.1  & 931.2
\end{tabular}
\caption{Gate parameters and total gate duration for the conditional (two-qubit) DDRF gates. The parameters are the RF frequency $\omega_{RF}/2\pi$, number of decoupling pulses $N$, RF voltage and interpulse delay $\tau$.} 
\label{table:con}
\end{table}

\begin{table}[h]
\begin{tabular}{c|c|c|c|c|c}

 & \shortstack{\phantom{g}$\omega_{RF}/2\pi$\phantom{g} \\ (kHz)} 
 & \shortstack{\phantom{g}$N$\phantom{g}\\ \phantom{(N)}}
 & \shortstack{RF voltage \\ (V)} 
 & \shortstack{\phantom{g}$\tau$ \phantom{g}\\ (\textmu s)} 
 & \shortstack{Total gate time \\ (\textmu s)} \\ \hline\hline
$n$   & 4173.388 & 16 & 0.319    & 37.62 & 1203.84 \\ \hline
$c_0$ & 2797.69  & 16 & 1.1849   & 15.2  & 486.4   \\ \hline
$c_1$ & 2754.55  & 16 & 0.8129   & 21.9  & 700.8   \\ \hline
$c_2$ & 2653.001 & 16 & 1.95     & 8.3   & 265.6   \\ \hline
$c_3$ & 2680.491 & 16 & 0.8767   & 14.5  & 464.0     \\ \hline
$c_4$ & 2676.095 & 16 & 0.478945 & 43.9  & 1404.8
\end{tabular}
\caption{Gate parameters and total gate duration for the unconditional (single-qubit) DDRF gates. The parameters are the RF frequency $\omega_{RF}/2\pi$, number of decoupling pulses $N$, RF voltage and interpulse delay $\tau$.}
\label{table:uncon}
\end{table}

\label{methods:gate_parameters}
Table \ref{table:con} (Table \ref{table:uncon}) shows the gate parameters and total gate duration for the optimized two-qubit (single-qubit) DDRF gates used in this work. The RF voltage is calibrated after the other parameters are chosen, by repeating the gate 22 or 34 times. Based on the coherent over- or underrotation error reported by two-qubit GST, the RF voltage is then adjusted before the gate is used in VQE. 

\subsection{Gate set tomography}
\label{methods:GST}
Gate set tomography requires a defined \emph{gate set} consisting of SPAM operations and gates. The SPAM operations correspond to initialization of the qubits in $|0\rangle$ and readout along Z, so in $|0\rangle$ or $|1\rangle$. The electron gates are defined as a XY4 decoupling sequence with Hermite shaped pulses and interpulse delay $\tau=12.96$ \textmu s, followed by a Hermite $\pi$/2 pulse \cite{Bartling2024}. To test the quality of the decoupling separately, we include an identity gate in the gate set, which is just the XY4 decoupling sequence. The nuclear-spin qubit gates are the gates that are optimized in this work and have a decoupling time $\tau$ tailored to minimize crosstalk. 

To implement GST we use the open-source python package PyGSTi \cite{Nielsen2020}. From specifications about the qubit processor, PyGSTi produces the gate sequences to run on the quantum register, that consist of preparation fiducials, germs and measurement fiducials. The fiducial gate blocks are used to prepare and readout in different bases, while the germs serve to amplify specific gate errors. The maximum germ depth for the two-qubit GST experiments was 16 gates. We run the sequences with 500 repetitions. To the experimental outcomes a Markovian model is fit using maximum-likelihood estimation. The model violation $N_{\sigma}$ indicates non-Markovianity in the system and is in our case between 36 and 64 standard deviations. To calculate the confidence intervals for the gate infidelities, we use the Hessian of the likelihood \cite{Nielsen2021}. We report confidence intervals of one standard deviation. Different error sources can be separated using the error generator formalism, like in Fig. \ref{fig:fig3}b. The error process of a gate $G$ can be separated from the ideal gate $\bar{G}$ using an error generator $L$, s.t. $G=\text{e}^L\bar{G}$. $L$ can now be linearly decomposed into elementary error generators \cite{Blume-Kohout2022}. The values in Fig. \ref{fig:fig3}b are calculated by applying only specific elementary error generators (part of the total error) to the ideal gate, and calculating the two-qubit average gate infidelity.

\subsection{VQE}
\label{methods:VQE}
VQE Hamiltonians were obtained using the OpenFermion python package \cite{McClean2020}. Molecular Hamiltonians were either calculated using the PySCF package \cite{Sun2017} via OpenFermion or were already present in the OpenFermion distribution. We used the sto-3g orbital set. For H$_2$, the full Hamiltonian is normally represented by 4 qubits, but by using a checksum code \cite{Steudtner2018}, the number of qubits can be reduced by 2, allowing the ground state to be encoded in only 2 qubits. For LiH, the full system consists of 4 electrons across 6 orbitals, which would require 12 qubits. We reduce the qubit requirements by assuming that the inner LiH shell is fully occupied (at the cost of the accuracy of the ground-state energy), removing 2 qubits, and further reduce the active space of the calculation to three orbitals that contribute the most to the ground-state energy. The remaining system requires 6 qubits to simulate, which is brought down to 4 by using the checksum code. At the bondlength, the ground-state energy of this approximated system is 2mHa higher than of the exact solution, sufficiently close not to be limiting for the presented experiments. 

\section*{Data availability}
All data underlying the study will be made available on the 4TU data server.
\section*{Code availability}
Code used to operate the experiments is available on request.

\putbib

\section*{Acknowledgements}
We thank H.P. Bartling, C.I. Ostrove and J. Gonzalez de Mendoza for valuable discussions. We gratefully acknowledge support from the joint research program “Modular quantum computers” by Fujitsu Limited and Delft University of Technology, co-funded by the Netherlands Enterprise Agency under project number PPS2007. 
This project has received funding from the European Research Council (ERC) under the European Union’s Horizon 2020 research and innovation programme (grant agreement No. 852410). This work was supported by the Dutch National Growth Fund (NGF), as part of the Quantum Delta NL programme. 
This work is part of the research programme NWA-ORC (NWA.1292.19.194), which is partly financed by the Dutch Research Council (NWO).
This project has received funding from the European Union’s Horizon Europe research and innovation program under grant agreement No 101135699.  
This research was supported by the education and training program of the Quantum Information Research Support Center, funded through the National research foundation of Korea (NRF) by the Ministry of Science and ICT (MSIT) of the Korean government (No. 2021M3H3A103657313).

The fitting algorithms were performed on the DelftBlue supercomputer at Delft University of Technology \cite{DHPC2022}.

\section*{Author contributions}
All authors devised the experiments. MvR, JY and HBvO prepared the experimental apparatus, performed the experiments and collected the data. All authors analyzed the data and wrote the manuscript. THT supervised the project.

\section*{Competing interests}
Delft University of Technology has been granted two patents, NL2037348 (inventors: M. Iuliano, A. Rodríguez-Pardo Montblanch, N. Demetriou, T.H. Taminiau, H.B. van Ommen, R. Hanson) and NL2038623 (inventors: C. Waas, N. Demetriou, H.B. van Ommen, R. Hanson, T.H. Taminiau, H.K.C. Beukers), that are related to DDRF gates.

\end{bibunit}

\clearpage

\onecolumngrid
\renewcommand{\thefigure}{S\arabic{figure}}
\setcounter{figure}{0}

\begin{center}
    {\Large \centering \bf Supplementary Information}
\end{center}
\section*{Contents}
\begingroup
\makeatletter
\def\@tocfile{supp.toc}   %
\@starttoc{supp}
\makeatother
\endgroup

\clearpage

\begin{bibunit}

\suppsection{Effective Rabi frequency}
\label{suppsec:eff-rabi}

\begin{figure*}
    \includegraphics[width=1 \textwidth]{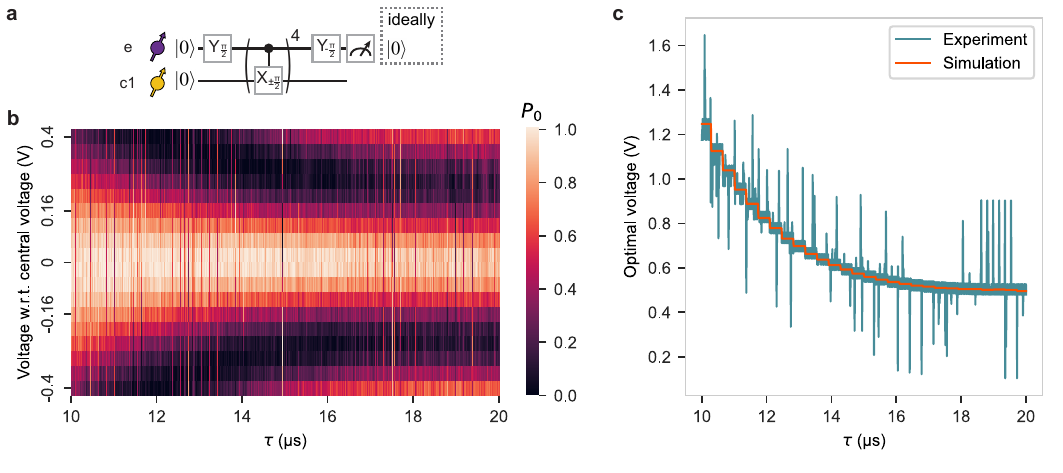}
    \caption{\emph{$\tau$-voltage sweep} a) Crosstalk measurement circuit. b) $P_0$ of the electron spin after four repetitions of the $c_1$ gate. While sweeping interpulse delay $\tau$, we change the center of the voltage sweep according to the effective Rabi frequency (Eq. \ref{eq:sup_eff_rabi_volt}). c) Comparison of the optimal voltage according to experiment and simulation.}
    \label{figs:tau-voltage_sweep}
\end{figure*}

To obtain data like in Fig. 2a of the main text, we sweep the interpulse delay $\tau$ of the DDRF gate under study. However, the Rabi frequency $\Omega$ (linearly dependent on the RF voltage $V_\text{RF}$) has to be adjusted to keep the rotation angle $\theta$ of the gate constant. The relation is not linear, but follows Eq. 3 of the main text. To confirm the accuracy of Eq. 3, we run the circuit shown in Fig. \ref{figs:tau-voltage_sweep}a, sweeping both $\tau$ and $V_{\text{RF}}$ of the two-qubit DDRF gate (Fig. \ref{figs:tau-voltage_sweep}b. The color indicates the electron-spin population in $\ket0$ ($P_0$) after the four applications of the gate. The vertical axis is the difference between the applied voltage $V_{\text{RF}}$ and the estimated optimal gate voltage $V_{\pi/2}(\tau)$. To estimate $V_{\pi/2}(\tau)$, we first calibrate a reference $\pi/2$ gate at $\tau_{\text{ref}}$, which gives the known-good voltage $V_{\text{ref}}=V_{\pi/2}(\tau_{\text{ref}})$. First, we approximate the actual RF driving time $\tau^{\text{RF}}$ using 
\begin{equation}
\label{eq:estimation}
    \tau^{\text{RF}} = \tau-t_{\text{buffer}}-2\times\frac{1}{2}t_{\text{roll}},
\end{equation}
where $t_{\text{buffer}} = 1.2$ \textmu s is a buffer time around the electron-spin qubit MW pulses and $t_{\text{roll}} = 4/f_{rf}$ is the duration of the $\sin^2$ roll-on and roll-off envelope of the RF pulse. Approximating the RF pulse as a square pulse, $V_{\pi/2}(\tau)$ can now be found using
\begin{equation}
\label{eq:sup_eff_rabi_volt}
    V_{\pi/2}(\tau) = V_{\text{ref}}\frac{\tau_{\text{ref}}^{\text{RF}}\beta(\tau_{\text{ref}}^{\text{RF}})}{\tau^{\text{RF}}\beta(\tau^{\text{RF}})},
\end{equation}
where $\beta(\tau) = \text{sinc}(\Delta_1\tau) \hspace{4pt}\mp\hspace{4pt}\text{sinc}(\Delta_0\tau)$ (i.e. Eq. 3).
Figure \ref{figs:tau-voltage_sweep}c shows a simulated required optimal voltage for the gate and the measured optimal value (the voltage that yields the highest electron-spin $P_0$ value in Fig. \ref{figs:tau-voltage_sweep}b.) The simulation calculates the Fourier transform of the RF pulses, because the bandwidth of the pulses determines the effective Rabi frequency \cite{VanOmmen2024}. The correspondence between the simulation and the data verifies the validity of our estimation described in Eq. \ref{eq:estimation} and the effective Rabi frequency described in Eq. 3 of the main text.

\suppsection{On-target gate infidelities}
\label{suppsec:on-target}
\begin{figure*}
    \includegraphics[width=1 \textwidth]{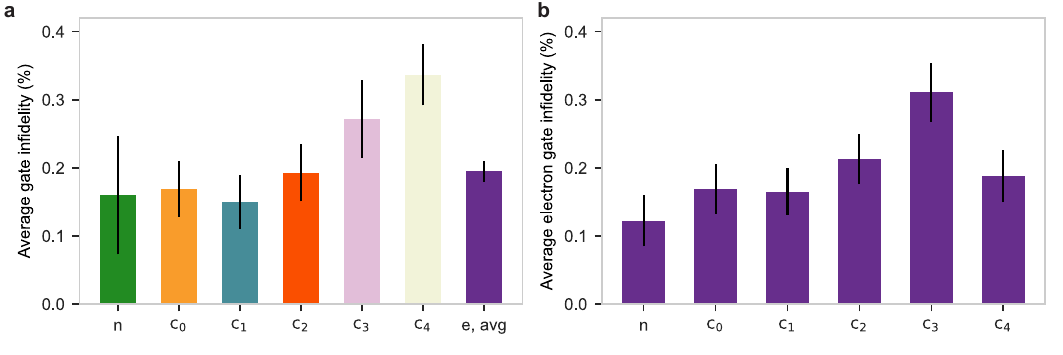}
    \caption{\emph{On-target gate infidelities} a) Average of the single-qubit gate infidelities in the single-qubit space of the target qubit. We average over the single-qubit X($\pi/2$) and Y($\pi/2$) rotations. b) For the electron-spin qubit, we obtain an average single-qubit gate infidelity from all six two-qubit GST experiments. The average is shown in a).}
    \label{figs:on_target}
\end{figure*}
The gate infidelities reported in Fig. 3 of the main text are defined in the two-qubit space. So even for a single-qubit gate, the infidelity due to errors on the other qubit is taken into account. This is in contrast to the metric that is measured in, for example, single-qubit randomized benchmarking, which is a single-qubit gate (in-)fidelity in the single-qubit space. However, it is possible to extract the `on-target' gate infidelity for the single-qubit gates from the GST two-qubit process matrices. These on-target infidelities are useful as they can be compared to single-qubit gate infidelities. We follow the approach reported by Madzik et al. (2022) \cite{Madzik2022}.

For the nuclear-spin qubit gates, we find an average fidelity of 99.79(2)\% (Fig. \ref{figs:on_target}a). This is an average over the six nuclear spins and over the single-qubit X($\pi/2$) and Y($\pi/2$) rotations. For the electron-spin qubit, we find an average of 99.81(2)\%. The electron-spin single-qubit gate fidelities are measured in all six two-qubit GST experiments (Fig. \ref{figs:on_target}b). In Fig. \ref{figs:on_target}a, for the electron-spin qubit, we report the average of the values in Fig. \ref{figs:on_target}b. 

We calculate the uncertainty on the on-target gate infidelities using the Hessian of the log-likelihood function, using pyGSTi \cite{Nielsen2020}.

\suppsection{Parametrized single-qubit gates}
\label{sec:parametrized_gates}
Optimizing a VQE circuit requires unitaries with arbitrary rotation axes and rotation angles. In this work, these arbitrary unitaries are prepared with parametrized single qubit gates. In this section, we discuss the construction of the parametrized gates for the electron spin and the nuclear spins. 

To implement an arbitrary unitary gate for the electron, we use two $\pi/2$-pulses around y with an XY4 decoupling sequence in between. Before the first y-pulse, before the XY4 and after the last y-pulse, we update the phase of the control microwave. This way, we effectively implement a $Z_{\theta_{1}}X_{\theta_{2}}Z_{\theta_{3}}$ gate that can produce any arbitrary single qubit unitary. For the electron-spin gates, this sequence was preferred over an amplitude-calibrated single pulse gate, because the electron is strongly coupled with the adjacent nitrogen nuclear spin, and therefore reducing the amplitude of the electron-spin pulse may start to introduce entanglement between the electron and the nitrogen nuclear spin.

For implementing an arbitrary nuclear-spin gate rotation angle, a fraction of the RF amplitude of the characterized $\pi/2$ gate, corresponding to the target rotation angle fraction of $\pi/2$, was used for the RF driving. A phase update on the RF oscillator was used to define an arbitrary rotation axis.

The LiH VQE circuit contains single-qubit gates around the z-axis. These are implemented by changing the reference frame.

\suppsection{VQE implementation details}
\label{suppsec:vqe-implementation}

\begin{figure*}
    \includegraphics[width=1 \textwidth]{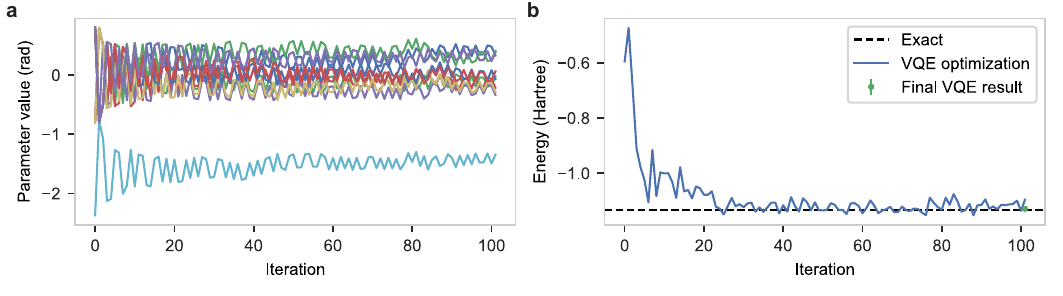}
    \caption{\emph{H$_2$ VQE optimization for interatomic spacing 0.8 Å} a) Circuit parameter evolution during the optimization. b) Ground state energy evolution during the optimization. }
    \label{figs:H2_vqe_fig}
\end{figure*}

This section describes experimental details on the implementation of VQE in this work. 
For H$_2$, the expectation value of the Hamiltonian is given by

\begin{equation}
    \langle H\rangle=C_0+C_1\langle Z_1I_2\rangle+C_2\langle I_1Z_2\rangle+C_3\langle Z_1Z_2\rangle+C_4\langle X_1X_2\rangle
\end{equation}
\\
where $C_i$ are constants which depend on the molecular properties and $\langle Z_1I_2\rangle$, $\langle I_1Z_2\rangle$, $\langle Z_1Z_2\rangle$ and $\langle X_1X_2\rangle$ are the expectation values of products of Pauli operators (see Methods E. for further details on how the form of $H$ is determined). For example, for two $H$ atoms separated by $0.74$ \si{\angstrom} (near the bond length), the constants are $C_0=-0.33832, C_1=C_2=0.394844, C_3=0.011246, C_4=0.181210$. The objective of VQE is to find the state $\ket{\psi}$ that minimizes $\langle H\rangle$. 

The state $\ket{\psi}$ is created by an ansatz quantum circuit, which for $H_2$ contains four single-qubit gates parameterized using the Euler angles, and an unparametrized two-qubit gate (for the implementation of the parametrized gates see section \ref{sec:parametrized_gates}). The phases of the first gates do not affect the measurement outcome as each qubits starts in $\ket{0}$, leaving a vector $\boldsymbol{\theta}$ containing $4\times3-2=10$ parameters to optimize over.

To estimate the $\langle H \rangle_{\boldsymbol{\theta}}$ for a specific ground state parametrized by $\boldsymbol{\theta}$ ($\ket{\psi}_{\boldsymbol{\theta}}$), the qubits are measured in the bases required to calculate the Pauli expectation values in $\langle H \rangle$. Commuting expectation values (for $H_2$: $Z_1I_2$, $ I_1Z_2$ and $Z_1Z_2$) can be sampled simultaneously be reading out each qubit individually. Although this grouping strategy gives rise to covariance between expectation values in the same group, causing a slightly larger variance on the total energy estimate, it has been shown that for a fixed number of total samples the variance on the energy estimate is much smaller when measurements are grouped this way \cite{Kandala2017}.

Before the first optimization iteration, we choose a $\boldsymbol{\theta}^0$ which yields the Hartree-Fock state (HF). The HF state is easy to calculate classically and ensures we start close to the ground state of the molecule. To optimize $\boldsymbol{\theta}$ to find the ground state, we use a gradient descent algorithm called simultaneous perturbation stochastic approximation (SPSA), implemented like in Kandala et al. \cite{Kandala2017}. It makes use of a perturbation size $c_k$ that decreases with iteration number $k$:

\begin{equation}
    c_k = \frac{c}{(1+Ck)^{\gamma}}
\end{equation}
\\
and a learning rate $a_k$:

\begin{equation}
    a_k = \frac{a}{(1+Ak)^{\alpha}}
\end{equation}
\\
Each optimization iteration $k$, each parameter $\theta_i$ is randomly chosen to increase or decrease by $c_k$, yielding a new set $\boldsymbol{\theta}_{k,+}$. In other words, a vector $\vec{s}_k$ is generated, with $s_{i,k}=\pm1$ at random, and $\boldsymbol{\theta}_{k,+} = \boldsymbol{\theta}_k + \vec{s}_k c_k$. The gradient in this direction is estimated by evaluating $\langle H \rangle_{\boldsymbol{\theta}_{k,+}}$ and $\langle H \rangle_{\boldsymbol{\theta}_{k,-}}$, where $\boldsymbol{\theta}_-$ is given by flipping all the signs of $\vec{s}_k$: $\boldsymbol{\theta}_{k,-} = \boldsymbol{\theta}_k - \vec{s}_k c_k$. The learning rate $a_k$ determines how much $\boldsymbol{\theta}_k$ is adjusted according to the gradient, yielding a new $\boldsymbol{\theta}_{k+1}$. We used $\gamma = 0.101$ and $\alpha = 0.602$, which are theoretically optimal values \cite{Spall1998}. For the other parameters we used $c = 0.4$, $C = 1$, $a = 0.8$, and $A = 1$, optimizing these values for a faster convergence of the ground state energy value is left for future research. 

To estimate $\langle H \rangle_{\boldsymbol{\theta}_{k,\pm}}$, we execute the quantum circuit for $\boldsymbol{\theta}_{k,\pm}$ two times (with 1000 shots each), once to obtain the expectation values for $Z_1I_2$, $ I_1Z_2$ and $Z_1Z_2$, and once to obtain the expectation values for $X_1X_2$. These circuits differ by their final parametrized gates, in which a rotation is absorbed that ensures readout of the correct Pauli. For example, a $-Y_{\frac{\pi}{2}}$ rotation gives an $X$ readout. 

After 100 trial states, we take the $\boldsymbol{\theta}_{k,\pm}$ that has the lowest average ground state energy and perform the circuit with $\boldsymbol{\theta}_{k}$ with 10000 shots to obtain a final estimation of the ground state energy. The total VQE experiment for one interatomic spacing takes approximately one hour. Every 30 minutes the magnetic field is calibrated. We make sure the calibration is never in between the measurement for one parameter set $\boldsymbol{\theta}$, to make sure the gradient estimation is reliable. 

For the results with the Clifford Fitting (CF, see section \ref{CF}), the expectation values for the products of Pauli operators are corrected based on the results of the CF, before they are fed into the SPSA algorithm. Figure \ref{figs:H2_vqe_fig} shows the optimization in which CF was used, for an interatomic spacing of 0.8 Å.

\begin{figure*}
    \includegraphics[width=1 \textwidth]{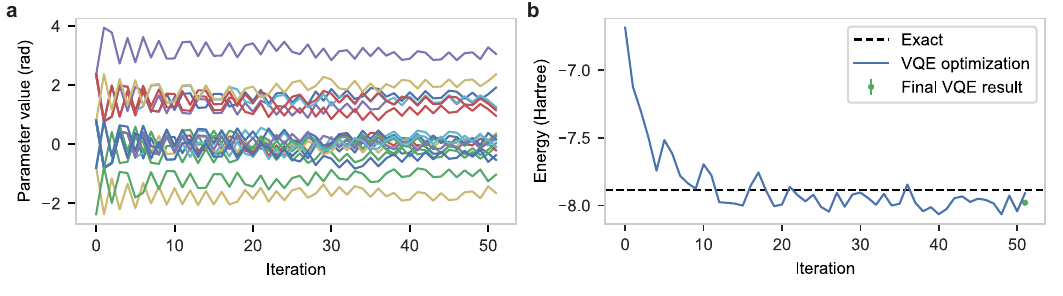}
    \caption{\emph{LiH VQE optimization for interatomic spacing 1.5 Å} a) Circuit parameter evolution during the optimization. b) Ground state energy evolution during the optimization. }
    \label{figs:LiH_vqe_fig}
\end{figure*}

VQE for LiH is implemented similarly to VQE for $H_2$. Like for $H_2$, to decrease the required number of experiments, Pauli expectation values of commutating terms in the Hamiltonian are extracted from the same measurement. For LiH we can calculate the Hamiltonian of 99 terms in 25 measurements, which we perform with 500 shots each. We measure 50 trial states in the optimization, after which the measurement of the final parameter set is done with 2000 shots. The total VQE experiment for one interatomic spacing takes 11 hours. Figure \ref{figs:LiH_vqe_fig} shows the optimization in which CF was used, for an interatomic spacing of 1.5 Å.

\suppsection{Clifford fitting} \label{CF} 

\begin{figure*}
    \includegraphics[width=1 \textwidth]{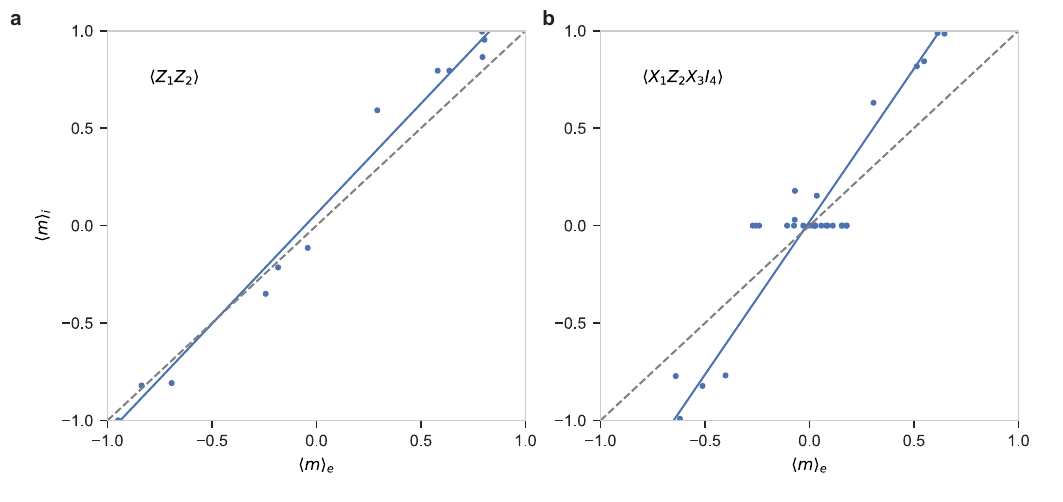}
    \caption{\emph{Clifford fitting comparing  the experimental values $\left<P\right>_e$ and the ideal values $\left<P\right>_i$.} a) Clifford fitting example for VQE of H$_2$, for the Pauli expectation value $\left<Z_1Z_2\right>$. b) Clifford fitting example for VQE of LiH, for the Pauli expectation value $\left<X_1Z_2X_3I_4\right>$. }
    \label{figs:cf}
\end{figure*}
VQE is a noisy intermediate-scale quantum (NISQ) application, where the expectation value of a quantum state is used and not the single shot result. For such applications, error mitigation can help improve the final outcome, in this case the estimate of $\langle H\rangle$ \cite{Cai2023, Guo2024}. In this work, we apply a method called Clifford Fitting (CF) that was first introduced in \cite{Czarnik2021}. A main advantage of using CF is that it is scalable in the sense that simulating Clifford gates can be done efficiently for larger circuits as well \cite{Bravyi2019}. In this section, we discuss the implementation details of CF.

The goal of CF is to find correction values per Pauli observable, based on learning of the noise in our system. We perform the following steps: 
\begin{enumerate}
    \item Choose additional parameter vector sets $\boldsymbol{\theta}^P_i$ for every Pauli observable $P$ in the Hamiltonian of the molecule, to construct circuits to learn the noise in our system.
    \item Extract two correction values per $P$ based on the effect of the system's noise on $P$, by comparing experimental and simulated executions of the circuits found using the first step.
    \item In every iteration of the VQE, correct the measured value for $\left<P\right>$ before computing $\left<H\right>$.
\end{enumerate}

Below, we elaborate on the different steps.

\suppsubsection{Parameter vector set selection}
The first step is to prepare multiple sets of parameter vectors $\boldsymbol{\theta}^P_i$ per Pauli operator $P$ in the Hamiltonian. We want to find sets of $\boldsymbol{\theta}^P_i$ that give spread out expectation values $\left<P\right> = [-1,1]$ for execution of the circuit. This way, when putting the ideal expectation value $\left<P\right>_i$ on the horizontal axis and the experimental expectation value $\left<P\right>_e$ on the vertical axis, a line can be fit to the points. In this work, we produce 500 random sets of $\boldsymbol{\theta}^P_i$. For H$_{2}$ (LiH), one set of $\boldsymbol{\theta}^P_i$ corresponds to $3\times4$ ($3\times8$) parameters to make 4 (8) unitaries $U_i(\boldsymbol{\theta}_i)$. Among the 4 (8) unitaries that we vary, 2 (6) unitaries are randomly chosen and constrained to be a random Clifford unitary. This ensures that the simulation of the circuit to get the ideal expectation value is scalable \cite{Guo2024, Bravyi2019}. Like in the actual VQE experiment, the start phase of the first 2 (4) gates are set to 0 since they do not affect the measurement outcomes. Among the 500 random sets of parameter vectors, for every $P$ we choose 10 parameter vector sets to run experimentally. 

First, we choose the parameter vector sets that gives the maximum and minimum value for each observable. Then we divide the expectation value range $R = \left<P\right>_{max}-\left<P\right>_{min}$ into 10 equal partitions.

Now we choose parameter vector sets that are closest to the next outer-most value $\left<P\right>_{max} - R/10$ and $\left<P\right>_{min} + R/10$. This provides that we have four close-to-extreme values for each observable. 

After this process, if we have enough samples, we stop the parameter set selection process. Otherwise, we continue with choosing the observable whose result is used the most in the Hamiltonian, and appending parameter sets that fills in the expectation value range by choosing values closest to $\left<P\right>_{min} + iR/10$, $i$ is integer in range [2,8].

\suppsubsection{Extraction of the correction values}
After the parameter sets are selected, the circuit belonging to each of the selected parameter vector sets is measured on the experimental hardware. The measurement result $\left<P\right>_e$ is plotted together with the ideal expectation value $\left<P\right>_i$. For ideal hardware, the data should be on a line $y=ax+b$ with $a = 1$ and $b = 0$. We fit the data to $y=ax+b$ and extract correction values $a_P$ and $b_P$. Figure \ref{figs:cf}a (\ref{figs:cf}b) shows an example for H$_2$ (LiH). 

\suppsubsection{VQE $\left<H\right>$ correction}
Running VQE, every iteration $k$, $\left<H\right>^k$ is calculated from the experimental Pauli expectation values $\left<P\right>^k_e$. The corrected value $\left<P\right>^k_{c}$ is calculated using $a_P$ and $b_P$:

\begin{equation}
    \left<P\right>^k_{c} = \frac{\left<P\right>^k_{e}-b_P}{a_P}.
\end{equation}

The CF VQE results in Figs. 4 and 5 of the main text are a result of this error mitigation process.

\suppsection{Initialization and readout}
\label{suppsec:init_and_ro}
In this section we elaborate on the initialization and readout required for the experiments in this work. 

\suppsubsection{SPAM implementation details}

\begin{figure*}
    \includegraphics[width=1 \textwidth]{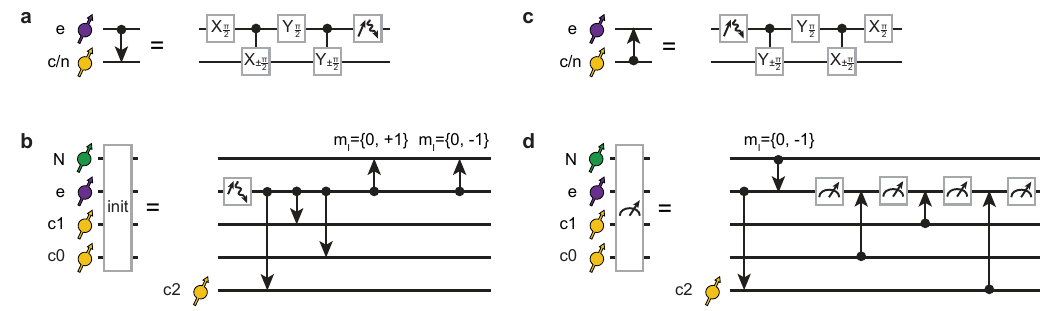}
    \caption{\emph{Initialization and readout} a) Reduced SWAP circuit for nuclear spin qubit initialization. b) Initialization circuit for LiH VQE. c) Circuit for nuclear spin qubit readout. d) Readout circuit for LiH VQE. }
    \label{figs:Init_ro_circuits}
\end{figure*}

The electron spin qubit is initialized in $m_s = 0$ using optical excitation of the $m_s=-1$ transition. The initialization of the nuclear spin qubits is done using a so-called reduced SWAP operation (Fig. \ref{figs:Init_ro_circuits}a), see also \cite{Bartling2024}.

Figure \ref{figs:Init_ro_circuits}b shows the initialization for the LiH VQE experiment. The nitrogen nuclear-spin qubit is a spin-1 system, so SWAP operations on the different subspaces are required. Note that in addition to the qubits that we use in the algorithm, we additionally initialize $c_2$, which we use for readout of the electron spin-qubit. 

For all the experiments except the LiH VQE, the electron is read out optically first where-after the inverse of the initialization circuit (Fig. \ref{figs:Init_ro_circuits}c) can be used to map the nuclear spin qubit state on the electron and readout the electron again. In all measurements, the electron readout is corrected for the measured SSRO infidelity, except in the GST experiments. GST takes the raw photon counts corresponding to the Z parity values of the two-qubit states.

For the LiH VQE, we could improve the contrast of the nitrogen nuclear spin qubit by not doing an optical readout of the electron first. This was achieved by using $c_2$ as an ancilla qubit, see Fig. \ref{figs:Init_ro_circuits}d.

\suppsubsection{SPAM fidelities reported by GST}

GST also obtains fidelities for SPAM. In Tab. \ref{table:SPAM}, we show the two-qubit state fidelity of the initial state and the fidelity of the two-qubit measurement POVM. Note that the nitrogen nuclear-spin qubit is initialized in the opposite state according to the GST analysis, explaining the near-zero fidelity.

\begin{table}[h]

\begin{tabular}{c|c|c}
 & \shortstack{Initial state fidelity \\ (\%)} 
 & \shortstack{Measurement fidelity \\ (\%)} \\ \hline\hline
$e-n$  & 2.6(2) & 86.10(9) \\ \hline
$e-c_0$ & 98.3(1) & 90.27(7) \\ \hline
$e-c_1$ & 99.2(1) & 89.48(7) \\ \hline
$e-c_2$ & 98.6(1) & 89.99(7) \\ \hline
$e-c_3$ & 94.7(2) & 86.13(9) \\ \hline
$e-c_4$ & 97.0(2) & 89.85(7)
\end{tabular}

\caption{Initial state fidelity and measurement fidelities from the 2Q GST reports.}
\label{table:SPAM}

\end{table}

\begin{figure*}
    \includegraphics[width=1 \textwidth]{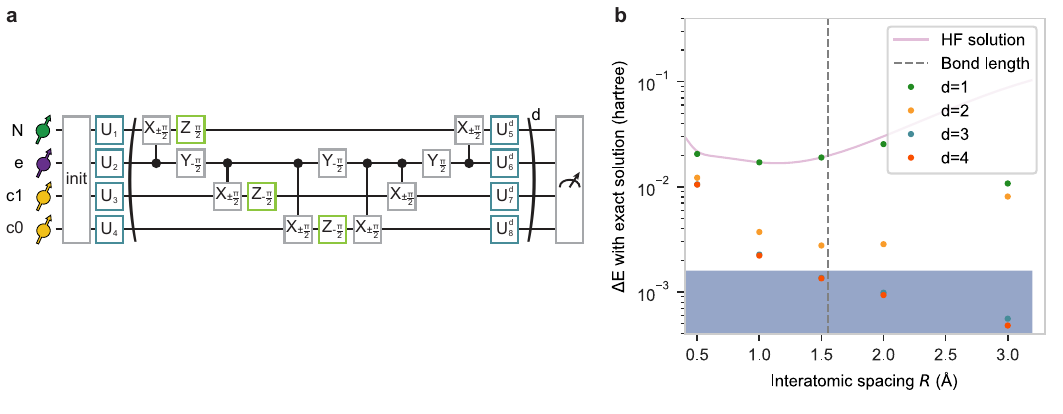}
    \caption{Simulation of LiH VQE for increasing circuit depth $d$. a) Simulated circuit. b) Difference with the exact solution for increasing circuit depth. `HF solution' is difference of the Hartree-Fock solution with the exact solution. The shaded region indicates chemical accuracy.}
    \label{figs:depth}
\end{figure*}

\begin{figure*}
    \includegraphics[width=1 \textwidth]{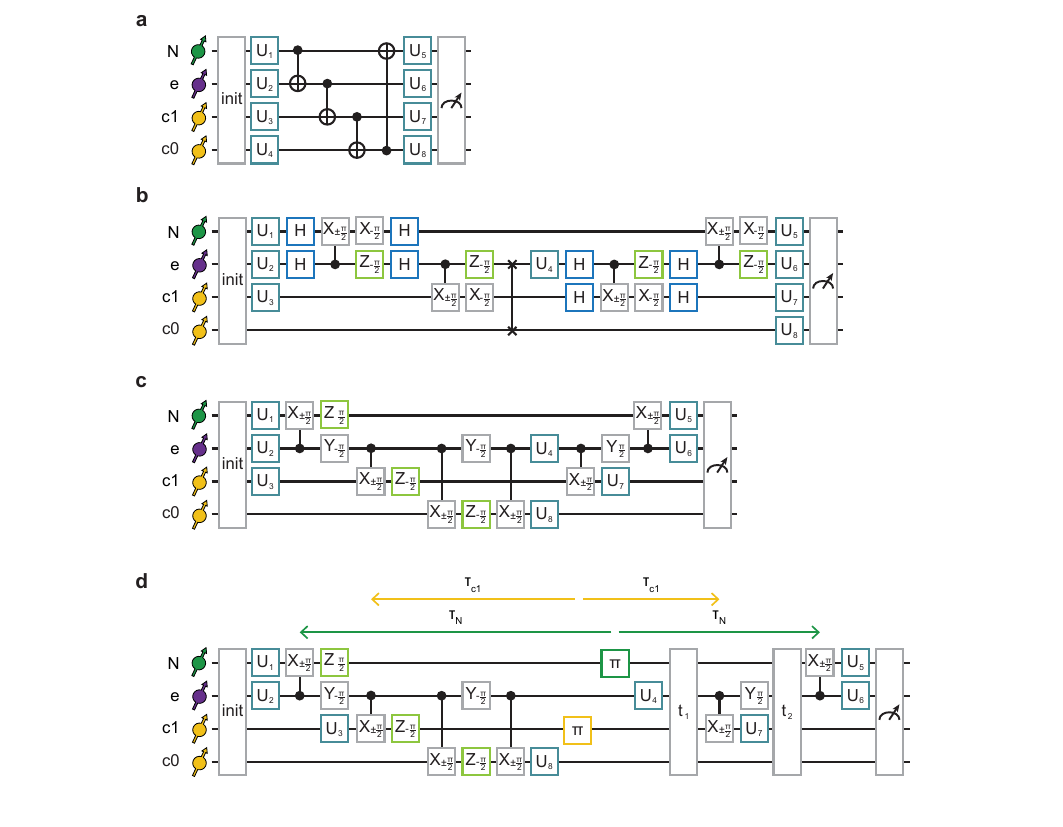}
    \caption{\emph{Circuit compilation for LiH} a) Hardware-agnostic starting point for the ansatz circuit. An entangling layer of cascading CNOTs is sandwiched between parametrized single-qubit gates. b) To minimize the amount of (long) nuclear spin gates, the state on qubit $c_0$ and the electron qubit are swapped. The CNOT gates are compiled into native gates. c) The SWAP gate is compiled and the amount of X and Y single qubit gates is reduced. d) The nitrogen qubit and $c_1$ qubit coherence is elongated by applying an echo pulse. }
    \label{figs:lih_circuit_compilation}
\end{figure*}

\suppsection{LiH VQE circuit depth analysis}
\label{suppsec:lih_depth}
The circuit in Fig. \ref{figs:lih_circuit_compilation}d is not deep enough to obtain chemical accuracy for LiH. In this section, we simulate higher-depth circuits for LiH, which are created by appending additional entangling layers and single qubit gates (Fig. \ref{figs:depth}a). We use Qiskit \cite{Javadi-Abhari2024} to construct the circuit. We start each simulation from the state that gives the HF state at the end of the circuit. Figure \ref{figs:depth}b shows the result up to depth 4. Depth 3 and depth 4 are sufficient to obtain chemical accuracy at the bond length of the molecule. 

\suppsection{LiH circuit optimization}
\label{suppsec:lih_circuit}
We optimized the ansatz circuit for LiH for our system. We start from an entangling layer of cascading CNOT gates, surrounded by single qubit gates with in total 8$\times$3-4=20 parameters (Fig. \ref{figs:lih_circuit_compilation}a). To minimize the total duration of the circuit, we utilize a SWAP gate. The SWAP changes around the role of $e$ and $c_0$, resulting in more (short) gates on the electron-spin qubit and less DDRF gates on $c_0$. Compilation of the CNOTs into native two-qubit gates then gives the circuit in Fig. \ref{figs:lih_circuit_compilation}b. Next, we compile the SWAP gate into native two-qubit gates. This only takes two two-qubit gates because c0 was in a known eigenstate before the application of the SWAP. As for the single qubit gates, the Z gates are virtual gates that can be implemented at no cost by changing the phase of the oscillator in the arbitrary waveform generator. Therefore, we change as many single qubit gates as possible into Z gates and simultaneously try to propagate single qubit gates to the beginning or the end of the circuit, where we can absorb them in the parametrized gates (\ref{figs:lih_circuit_compilation}c). 

Finally, we consider coherence times and echoes (Fig. \ref{figs:lih_circuit_compilation}d). The electron is already decoupled throughout the circuit. Qubit $c_0$ is in an eigenstate, then undergoes some gates with no waiting time in between. After $U_8$, we only care about the z projection. This means $c_0$ does not need an echo pulse. For the nitrogen spin qubit and $c_1$, we do implement echoes, which are DDRF gates. The echoes are placed such, that they are symmetric in time in between the two-qubit gates on the respective qubits. To achieve this, we add waiting times $t_{n}$ and $t_{c1}$ to the circuit.

\clearpage

\putbib

\end{bibunit}
\clearpage

\end{document}